\documentstyle[preprint,aps,epsf]{revtex}

\begin{document}

\title{
{\normalsize\hfill Kyoto University, Department of Astronomy, 98-38}\\
{\normalsize\hfill Osaka University, OUTAP-88}\\~\\
Time scales of relaxation and Lyapunov instabilities in
a one-dimensional gravitating sheet system}

\author{Toshio Tsuchiya}
\address{Department of Astronomy, Kyoto University, Kyoto, 606-8502, Japan}

\author{Naoteru Gouda}
\address{Department of Earth and Space Science, Osaka University,
Toyonaka, 560-0043, Japan}

\maketitle

\begin{abstract}
The relation between relaxation, the time scale of Lyapunov instabilities,
and the Kolmogorov-Sinai time in a one-dimensional gravitating sheet system
is studied. Both the maximum Lyapunov exponent and the Kolmogorov-Sinai
entropy decrease as proportional to $N^{-1/5}$. The time scales determined
by these quantities evidently differ from any type of relaxation time found
in the previous investigations. The relaxation time to quasiequilibria
(microscopic relaxation) is found to coincide with the inverse of the
minimum positive Lyapunov exponent. The relaxation time to the final thermal
equilibrium differs to the inverse of the Lyapunov exponents and the
Kolmogorov-Sinai time.
\end{abstract}

\narrowtext

\section{Introduction}

Relaxation is the most fundamental process in evolution of many-body
system. The classical statistical theory is based on ergodic property, which 
is considered to be established after relaxation. However, not all systems
do not show such an idealistic relaxation. A historical example is FPU
(Fermi-Pasta-Ulam) problem\cite{ferm65}, which experiences the induction
phenomenon (e.g., Ref\cite{hiro69,saito-n70}) and does not relax to the
equipartition for very long time.

From nearly thirty years of investigation, one-dimensional self-gravitating
sheet systems (OGS) have been known by their strange behavior in
evolution. Hohl\cite{hohl67a,hohl67b,hohl68} first asserted that OGS relaxes
to the thermodynamical equilibrium (the isothermal distribution) in a time
scale of about $N^2\,t_c$, where $N$ is the number of sheets, and $t_c$ is
typical time for a sheet to cross the system. Later, more precise numerical
experiments determined that the Hohl's result was not right, and then
arguments for the relaxation time arose in 1980's. A Belgian
group\cite{luwe84,seve84a} claimed the OGS relaxed in shorter than $Nt_c$,
whereas a Texas group\cite{wrig82,miller-b90} showed that the system showed
long lived correlation and never relaxed even after $2N^2t_c$. Tsuchiya,
Gouda and Konishi (1996)\cite{tsuc96a} suggested that this contradiction can
be resolved in the view of two different types of relaxations: the {\em
microscopic} and the {\em macroscopic} relaxations. At the time scale of
$Nt_c$, cumulative effect of the mean field fluctuation makes the energies
of the individual particles change noticeably. Averaging this change gives
the equipartition of energies, thus there is a relaxation at this time
scale. By this relaxation the system is led not to the thermal equilibrium
but only to a quasiequilibrium. The global shape of the one-body
distribution remains different from that of the thermal equilibrium. This
relaxation appears only in the microscopic dynamics, thus it is called the
microscopic relaxation. The global shape of the one-body distribution
transforms in much longer time scale. For example, a quasiequilibrium (the
water-bag distribution, which has the longest life time) begins to transform
at $4\times10^4Nt_c$ in average. Tsuchiya et al.(1996) \cite{tsuc96a} called
this transformation the macroscopic relaxation, but later in Tsuchiya, Gouda
and Konishi (1998) \cite{tsuc98c}, it is shown that this transformation is
onset of the {\em itinerant stage}. In this stage, the one-body distribution
stays in a quasiequilibrium for some time and then changes to other
quasiequilibrium. This transformation continues forever. Probability density
of the life time of the quasiequilibria has a power law distribution with a
long time cut-off and the longest life time is $sim 10^4Nt_c$.  Only by
averaging over a time longer than the longest life time of the
quasiequilibria, the one-body distribution becomes that of the thermal
equilibrium, which is defined as the maximum entropy state. Therefore the
time $\sim 10^6Nt_c$ is necessary for relaxation to the thermal equilibrium,
and called the {\em thermal relaxation time}. Although there are some
attempts to clarify the mechanisms of these relaxations
\cite{tsuc96a,miller-b96,milano98,tsuc98c}, the reason why the system does
not relax for such a long time is still unclear.

At the view of chaotic theory of dynamical systems, relaxation is understood
as mixing in phase space, and its time scale is given by the
Kolmogorov-Sinai time (KS time), $\tau_{\rm KS}=1/h_{\rm KS}$, where $h_{\rm
KS}$ is the Kolmogorov-Sinai entropy. However, it does not simply correspond
to the relaxation of the one-body distribution function, which is of
interest in many-body systems. Recently, Dellago and Posch\cite{dell97}
showed that in a hard sphere gas, the KS time equals the mixing time of
neighboring orbits in the phase space, whereas the relaxation of the
one-body distribution function corresponds to the collision time between
particles. Now, it is fruitful to study relation between relaxation and some
dynamical quantities, such as the KS entropy and the Lyapunov exponents, in
the OGS. Milanovi\'{c} et al.\cite{milano98} showed the Lyapunov spectrum
and the Kolmogorov-Sinai entropy in the OGS for $10\leq N \leq 24$. However,
since it is known that the chaotic behavior changes for $N\sim30$ for the
OGS\cite{reid93}, it is considerably important to extend the analysis to the
system larger than $N\sim30$. In this paper, we extend the number of sheets
to $N=256$ and follow the evolution numerically up to $T\sim 10^6Nt_c$, which 
is long enough for the thermal relaxation \cite{tsuc98c}.

\section{Numerical simulations}

The OGS comprises $N$ identical plane-parallel mass sheets, each of which
has uniform mass density and infinite in, say, the $y$ and $z$
direction. They move only in the $x$ direction under their mutual
gravity. When two of the sheets intersect, they pass through each other. The 
Hamiltonian of the system has the form
\begin{equation}
H=\frac{m}{2}\sum_{i=1}^N v_i^2  + (2\pi Gm^2)\sum_{i<j} |x_j-x_i|,
\end{equation}
where $m$, $v_i$, and $x_i$ are the mass (surface density), velocity,
and position of the $i$th sheet, respectively. Since the
gravitational field is uniform, the individual particles moves
parabolically, until they intersect with the neighbors. Thus the
evolution of the system can be followed by solving quadratic equations. This 
property helps us to calculate long time evolution with a high
accuracy. Since length and velocity (thus also energy) can be scaled in the
system, the number of the sheets $N$ is the only free parameter. The
crossing time is defined by
\begin{equation}
t_c=(1/4\pi GM)(4E/M)^{1/2},
\end{equation}
where $M$ and $E$ is the total mass and total energy of the system.
Detailed descriptions of the evolution of the OGS can be found in our
previous papers\cite{tsuc94b,tsuc96a,tsuc98c}.

In order to investigate dynamical aspects of the system, we calculated the
Lyapunov spectrum. The basic numerical algorithm follows Shimada and
Nagashima\cite{shimada-i79}, and detailed description of the procedure for
the OGS can be found in ref\cite{tsuc94b,milano98}. We made numerical
integration for $8\leq N \leq 128$ up to $10^8t_c$, which is enough time for
the system to relax, and up to $1.8\times10^7t_c$ for $N=256$ for reference.

\section{Results}

\begin{figure}[h]
\epsfbox{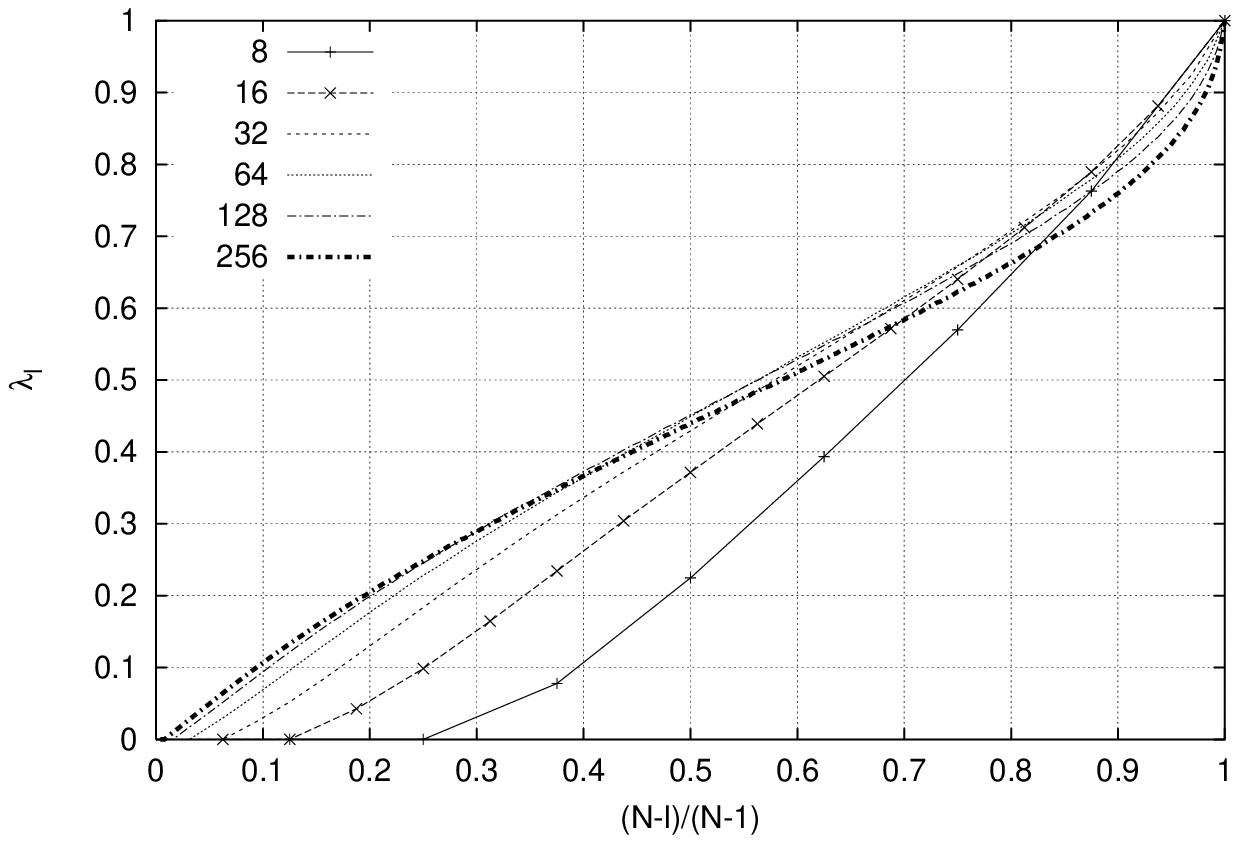}
\vspace{5mm}
\caption{Spectrum of the positive Lyapunov exponents for various $N$. The
index of the Lyapunov exponents is scaled to 0 to 1.0. The vertical axis
shows the Lyapunov exponents normalized by the value of the maximum Lyapunov 
exponent.
}
\label{fig:spectrum}
\end{figure}

Figure \ref{fig:spectrum} shows the spectrum of the Lyapunov exponents,
$\{\lambda_i\}$, where their unit is $1/t_c$. This figure is the same
diagram as Fig. 6 in Milanovi\'{c} et al\cite{milano98}, but the range of
$N$ is extended to $8\leq N \leq 256$. In the horizontal axis, $l$ is the
index of the Lyapunov exponents, which is labeled in the order from the
maximum to the minimum. Thus all the positive Lyapunov exponents $(l\leq N)$
is scaled between 0 to 1 in the axis. The vertical axis shows the Lyapunov
exponents normalized by the maximum Lyapunov exponents,
$\lambda_1$. Milanovi\'{c} et al\cite{milano98} stated that the shape of the
spectrum approximately converges for large $N$.A closer look, however, shows
bending of the spectrum, which is most clearly seen at $(N-l)/(N-1)\sim0.9$. 
This bending seems increase with $N$ for $N\geq32$. A further investigation
is needed to give a definite conclusion about the convergence of the shape
of the spectrum.

\begin{figure}[h]
\epsfbox{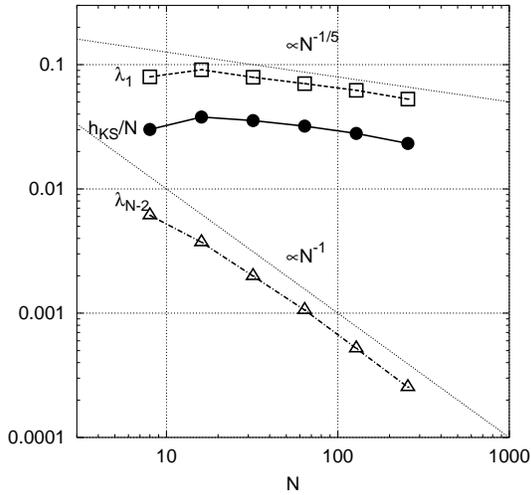}
\vspace{5mm}
\caption{Dependence of the KS entropy (solid line with the symbol
$\bullet$), the maximum Lyapunov exponent (long dashed curve with the symbol
$\Box$), and the minimum positive Lyapunov exponent (dashed dotted curve
with the symbol $\triangle$).}
\label{fig:Ndependence}
\end{figure}

Figure \ref{fig:Ndependence} shows $N$-dependence of the maximum
$(\lambda_1)$, the minimum positive Lyapunov exponent $(\lambda_{N-2})$, and
the KS entropy $h_{KS}$ per the number of freedom. $\lambda_1$ is already
shown in Fig.13 in Tsuchiya et al.\cite{tsuc94b}, and it is proportional to
$N^{-1/5}$ for $N\geq32$. Decreasing nature of the Lyapunov exponent may
indicate that the OGS approaches closer to an integrable system for larger
$N$.

As expected from the spectrum the KS entropy divided by $N$ is also
proportional to $N^{-1/5}$. Therefore the conjecture by Benettin et
al.\cite{bene79} that $h_{KS}$ increases linearly with $N$ is not right.
It is clear that the inverses of both the maximum Lyapunove exponents and
the KS entropy do not give the time scale of any type of relaxation time.

The $N$-dependence of small positive Lyapunov exponents are quite different
from larger ones. In Fig.\ref{fig:Ndependence}, the minimum positive
Lyapunov exponent, $\lambda_{N-2}$, is shown by a dashed dotted line with
the symbol $\triangle$. It decreases linearly for $N\geq32$, and its time
scale $1/\lambda_{N-2}$ is about the same as the microscopic relaxation
time ($\sim Nt_c$).

\begin{figure}
\epsfbox{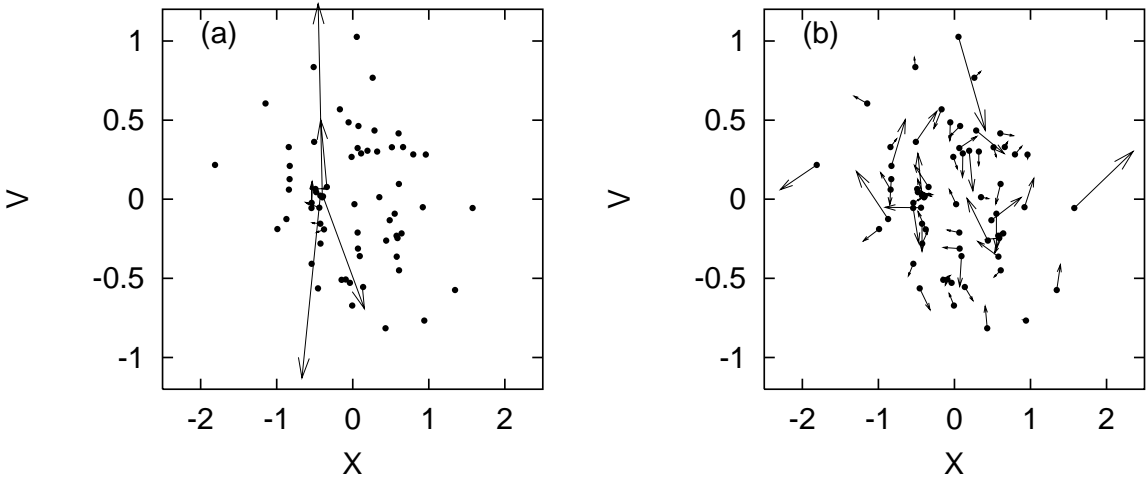}
\caption{Lyapunov eigen vectors for $N=64$. Filled circles indicate
positions of $N$ sheets and the arrows give direction of the Lyapunov eigen
vector: (a) the eigen vector for $\lambda_1$, (b) that for $\lambda_{N-2}$.}
\label{fig:LyapVec}
\end{figure}

The eigen vectors for the Lyapunov exponents also give a useful information. 
Figure \ref{fig:LyapVec} shows projection of the eigen vector for $N=64$ on
to the one-body phase space. Filled circles indicate positions of $N$ sheets
at a moment and the arrows give the direction of the Lyapunov eigen vector,
which grows with the rate of the Lyapunov exponent. Fig \ref{fig:LyapVec}(a)
is for the maximum Lyapunov exponent $\lambda_1$, and
Fig.\ref{fig:LyapVec}(b) is for the minimum positive one, $\lambda_{N-2}$.
For $\lambda_1$, the instability is carried only by a few particles, which
are interacting in a very small region. The instability is thus not for
global transformation. On the other hand, the instability with
$\lambda_{N-2}$ makes all particles mix in the phase space. This is the very
effect of relaxation. These features are commonly seen for different $N$.

The results that the coincidence of the $1/\lambda_{N-2}$ and the
microscopic relaxation time, and the direction of the eigen vector,
 may be suggesting that the microscopic relaxation time is
determined by the growing time of the weakest instability, which is
determined by the minimum positive Lyapunov exponent; in other words,
this time is necessary for the phase space orbit to mix in the phase space
in the all directions of freedom. In our working model of the evolution of
the OGS\cite{tsuc96a,tsuc98c}, the phase space is derived by some barriers
which keep the phase orbit inside for a long time. The microscopic
relaxation is considered to be a diffusion process in the barierred
region\cite{tsuc96a,miller-b96}, and in the time $\sim Nt_c$, restricted
ergodicity is established within the barierred region. This time may
correspond to the diffusion time in the slowest direction.

\section{Conclusions and Discussion}

In the ergodic theory, the KS time represents the time scale of ``Mixing''
in the phase space. On the other hand, the relaxation of the one-body
distribution is of the most interest in systems with large degrees of
freedom. We have shown that the time scale of the relaxation of one-body
distribution (both the microscopic and thermal relaxation) is certainly
different from that of the KS time, and found that the growing time of the
weakest Lyapunov instability is about the same as the microscopic relaxation
time. In addition, taking into account the direction of the eigen vector of
the weakest Lyapunov exponent, it is suggested that the microscopic
relaxation is determined by the weakest Lyapunov instability.

The KS entropy is defined as a typical time for the system to increase
``information''. This definition does not depend on the number of degrees of
freedom. In higher dimensions, however, even very small growth of
instability can increase information quite rapidly.  Therefore the KS time
does not seem suitable to characterize the relaxation of the one-body
distribution function.

The relaxation of the one-body distribution function implies ergodicity. To
attain ergodicity, the phase space orbits should diffuse over all accessible
phase space. For the microscopic relaxation, even though it is not true
thermal relaxation, the system shows ergodicity which is restricted in a
part of the phase space\cite{tsuc96a}. Therefore it seems natural that the
microscopic relaxation time is characterized by the slowest time of
diffusion. That may be the reason that the inverse of the minimum positive
Lyapunov exponent coincides the microscopic relaxation time.

Gurzadyan and Savvidy\cite{gurz86} derived the KS time in the usual three
dimensional stellar systems, which was found to be proportional to
$N^{1/3}$, by the method of the geodesic deviation. They asserted that the
system relaxes in this time scale. This result was not supported by
numerical simulations\cite{saka91} and by a semi-analytical
study\cite{good93b}. Our analysis of one-dimensional systems also gives
negative result for the conjecture of Gurzadyan and Savvidy. According to
our results, the time scale of the minimum positive Lyapunov exponent would
be related to the relaxation also in three-dimensional systems, though it is
difficult to show it numerically.

We have found that the KS time and any of the times of Lyapunov
instabilities do not give the thermal relaxation time in the OGS. In our
working model, the thermal relaxation is the successive transitions of the
phase space orbit among the barierred regions. The each region has locally
restricted ergodicity, hence the long time average could only give average
of the Lyapunov exponents, which are defined in each region, and it is
reasonable that the averaged Lyapunov exponents do not characterize the
thermal relaxation. Actual time of the thermal relaxation is that of
transition among quasiequilibria. It is necessary to find appropriate
dynamical quantities to describe it.

\acknowledgements
The authors are grateful to T. Konishi for many critical suggestions. We
also thank B. N. Miller and Y. Aizawa for valuable discussions. This work
was supported in part by JSPS Research Fellow and in part by the
Grant-in-Aid for Scientific Research(No.10640229)from the Ministry of
Education, Science, Sports and Culture of Japan.



\epsfxsize=10cm

\epsfxsize=10cm

\epsfxsize=16cm

\end{document}